\documentclass[aps,prb,twocolumn,groupedaddress,showpacs]{revtex4}
\usepackage{graphicx}

\begin{document}


\title{On negative streamers: a deterministic approach}


\author{Manuel Array\'as}
\affiliation{Universidad Rey Juan Carlos, Dept. de F\'{\i}sica, Tulip\'an s/n, 28933, M\'ostoles, Madrid, Spain.}


\date{\today}

\begin{abstract}
In dielectric breakdown, the phenomena of streamers formation and spontaneous branching is commonly observed. A deterministic negative streamer model is reviewed in this article. We recently have proposed that this reduced model is able to explain the branching phenomena, due to a Laplacian instability quite like in viscous fingering. Particular attention is paid to the physical interpretation of the model.  
\end{abstract}

\pacs{52.80.Mg, 47.54.+r, 51.50.+v, 05.45.-a}

\maketitle
\section{INTRODUCTION}
Atoms of any substance are electrically neutral. If we apply an electric field to a volume filled with neutral particles, the electric current will not flow through that volume. Since no charged particles are present, there will not be any electric current, which is in fact the directed motion of charged particles. Thus, the volume, filled with atoms of any substance, is an almost ideal insulator. 

The air is a good example of such an insulator. Every cubic centimetre of air contains roughly speaking $2.7 \times 10^{19}$ molecules of oxygen (O$_2$), nitrogen (N$_2$), vapour (H$_2$O) and some other gases. Every atom of oxygen contains 8 positively charged protons and the same number of negatively charged electrons. Every atom of nitrogen contains 7 protons and 7 electrons. It might seem there are more than enough charged particles, but those particles are bound by powerful electric forces to form electrically neutral atoms and molecules, and as a result of which the air is a perfect insulator. 

However, if a strong electric field is applied to matter of low conductivity and some electrons or ions are created by some agent, then these few mobile charges can generate an avalanche of more charges by impact ionization. A low temperature plasma is being created, resulting in an electric discharge. Examples range from natural phenomena like the familiar lightning and St.Elmo's fire to lamps (neon tubes, hight brightness flat computer and TV screens) and industrial plasma reactors for combustion gas cleaning, ozone generation, etc.\cite{Eddie} There have been a huge development of technologies based on this phenomena. Laser pumping and ion chambers used for particle detectors are other examples one can find. 

On the other hand, the understanding of the basic mechanisms of the electric discharge is a challenging problem where ideas from nonequilibrium thermodynamics, atomic physics, electromagnetism and pattern formation come into play. Discharges are nonequilibrium processes occurring in initially nonionized matter exposed to strong electric field. Depending on the spatiotemporal characteristics of the electric field and on the ionization and charge transport properties of the medium, discharges can assume many different modes of appearance. Phenomenologically discharges can be classified in stationary ones, such as arc, glow or dark discharges and transient ones, such as leaders, initial stages of sparks and streamers.\cite{Rai} As a warning, the distinction between the various discharge phenomena seems to vary among authors. 

A streamer is a sharp nonlinear ionization wave propagating into a nonionized gas, leaving a nonequilibrium plasma behind. They have been reported to appear in early stages of atmospheric discharges \cite{Pasko} like sparks or sprite discharges.\cite{Web} It is observed that streamers can branch spontaneously, but how this branching is precisely determined by the underlying physics is essentially an open question. We have proposed in recent work \cite{ME} a branching mechanism which is qualitatively different from other ideas. The older concept of dielectric breakdown can be traced back to Raether's work.\cite{Raether} In his model he introduced the rare long ranged photo-ionization events. Some stochastic models for dielectric breakdown have been proposed and studied since then.\cite{Niemeyer} We used a fully deterministic model with pure impact ionization and it becomes a surprise streamers get unstable and develop branching. The mechanism for this branching is related to a Laplacian interfacial instability. 

In this paper we will start introducing a basic fluid model incorporating the physical ingredients to describe a nonattaching gas such as nitrogen under normal conditions. In Sec.~III some numerical simulations based on this model are shown. In Sec.~IV and the ones to follow, the streamers branching is investigated by analytical means. We summarise the main results for stationary planar fronts and in Sec.~V for shock fronts. In Sec.~VI we set up the framework of the linear perturbation analysis for transversal Fourier modes, first the equation of motion and then the boundary conditions and the solution strategy. In Sec.~VII we obtain the asymptotic behaviour of the dispersion relation and finally, we end with a summary and prospect of future work.

\section{THE STREAMER MODEL}
In this section we present a {\it minimal streamer model}, i.e., a ``fluid approximation'' with local field-dependent impact ionization reaction 
in a nonattaching gas like argon or nitrogen.\cite{Ute} It treats the dynamics of the free electrons and positive ions in a homogeneous gas at rest.  In detail, this is as follows:

\noindent
({\it i}) The ionization by electrons in the gas is essentially the primary process in all spark discharge. Initially, an electron liberated by any  outside agents, as e.g radiation, is accelerated in a strong local field. It collides into a neutral molecule and ionises it. The result is a generation of new free electrons and a positive ion. In general, this process is determined by the rate of gain of energy of the electrons and their ability to produce ionization once they have sufficient energy. The energy gain depends on the field strength and the free path of electrons.  The free path depends on the pressure and character of the gas. The calculation from first principles of all this process is not an easy task. Fortunately, one can measure it quite simply by a procedure developed by Townsend.\cite{Loeb} We can then obtain the effective cross-section $\alpha_0$ for a given external $E_0$ electric field, and use Townsend approximation $\alpha_0\; \alpha(|{\cal E}|/E_0)=\alpha_0\;\exp(-E_0/|{\cal E}|)$ to write the following balance equations for electrons and ions

\begin{eqnarray}
  \label{eq:balance}
 \partial_\tau n_e+\nabla_{\bf R}\cdot{\bf j}_e=|\mu_e{\cal E}n_e|\;\alpha_0\;\alpha(E_0/|{\cal E}|)\\
 \partial_\tau n_i+\nabla_{\bf R}\cdot{\bf j}_i=|\mu_e{\cal E}n_e|\;\alpha_0\;\alpha(E_0/|{\cal E}|)
\end{eqnarray}
\noindent
where $n_{e,i}$ and ${\bf j}_{e,i}$ are particle densities and currents
of electrons and ions, respectively, and ${\cal E}$ is the electric field.
The fact that the source terms at the right hand side of the equations are equal is due to charge conservation in an ionization event. 

\noindent
({\it ii}) The electron particle current ${\bf j}_e$ is approximated as the sum of a drift and diffusion term
\begin{equation}
  \label{eq:current}
 {\bf j}_e=-\mu_e{\cal E}n_e-D_e\nabla_{\bf R}n_e 
\end{equation}
where $\mu_e$ and $D_e$ are the mobility and diffusion coefficient of the electrons. For anode-directed streamers the ion current can be neglected because it is more than two orders of magnitude smaller that electronic one, so we will take 
\begin{equation}
\label{eq:icurrent}
{\bf j}_i=0.
\end{equation}

\noindent
({\it iii}) The modification of the externally applied electric field 
through the space charges of the particles according to the Poisson equation
\begin{equation}
  \label{eq:poisson}
 \nabla_{\bf R}\cdot{\cal E}=e(n_i-n_e)/\varepsilon_0.
\end{equation}
It is this coupling between space charges and electric field 
which makes the problem nonlinear.

We want to add a few remarks. In the source term, ionization due to the photons created in recombination or scattering evens is neglected. This can be justified if the cross section of the photoionization process is much smaller than that due to electrons. Note that photoionization can be taken into account, but the dynamical equations will become nonlocal. In attaching gases like oxygen, a third kind of charged species needs to be taken into account, namely negative ions formed by a neutral molecule catching a free electron.\cite{Dhali} The equations are deterministic and stochastic effects are not accounted for in the model. 

Finally, the model must be complemented with appropriate boundary and initial conditions. Boundary conditions will be discussed in detail in the next section. For initial conditions, we ignore details of the plasma nucleation event (e.g. triggering by radiation from an external source), and assume that at $t=0$ a small well-localized ionization seed is present. We also make it clearer below. 

In order to identify the physical scales and the intrinsic parameters of the model, it is convenient to reduce the equations to dimensionless form. The natural units of the model are given by the ionization length 
$R_0=\alpha_0^{-1}$, the characteristic impact ionization field $E_0$ and
the electron mobility $\mu_e$, determining the velocity $v_0=\mu_e E_0$
and the time scale $\tau_0=R_0/v_0$. The values of those quantities for nitrogen at normal conditions are 
\begin{eqnarray}
  \label{eq:values}
  \alpha_0^{-1} \approx 2.3\; \mu\mathrm m, \;\;\;  E_0  \approx 200 \;\mathrm {kV/m}, \;\;\; \mu_e \approx 380 \;\mathrm {cm^2/Vs}. \nonumber
\end{eqnarray}
Hence we introduce the dimensionless coordinates \cite{Ute} 
${\bf r}={\bf R}/R_0$ and $t=\tau/\tau_0$, the dimensionless field
${\bf E}={\bf {\cal E}}/E_0$, the dimensionless electron
and ion particle densities $\sigma=n_e/n_0$ and $\rho=n_i/n_0$ 
with  $n_0=\varepsilon_0 E_0/(e R_0)$,
and the dimensionless diffusion constant $D=D_e/(R_0v_0)$. 

After this rescaling, the model reads
\begin{eqnarray}
\label{1}
\partial_t\;\sigma \;-\; 
\nabla\cdot {\bf j}
&=& \sigma \; f(|{\bf E}|)~,
\\
\label{2}
\partial_t\;\rho \;
&=& \sigma \; f(|{\bf E}|)~,
\\
\label{3}
\rho - \sigma &=& \nabla\cdot{\bf E}~,
\\
\label{4}
\sigma\;{\bf E} + D\;\nabla\sigma &=& {\bf j}~.
\end{eqnarray}
The function $f(|{\bf E}|)$ due to Townsend's expression yields
\begin{equation}
\label{ft}
f(|{\bf E}|)=|{\bf E}|\;\alpha(|{\bf E}|)= |{\bf E}|exp(-1/|{\bf E}|)
\end{equation}

\begin{figure*}
\includegraphics[width=0.24\linewidth]{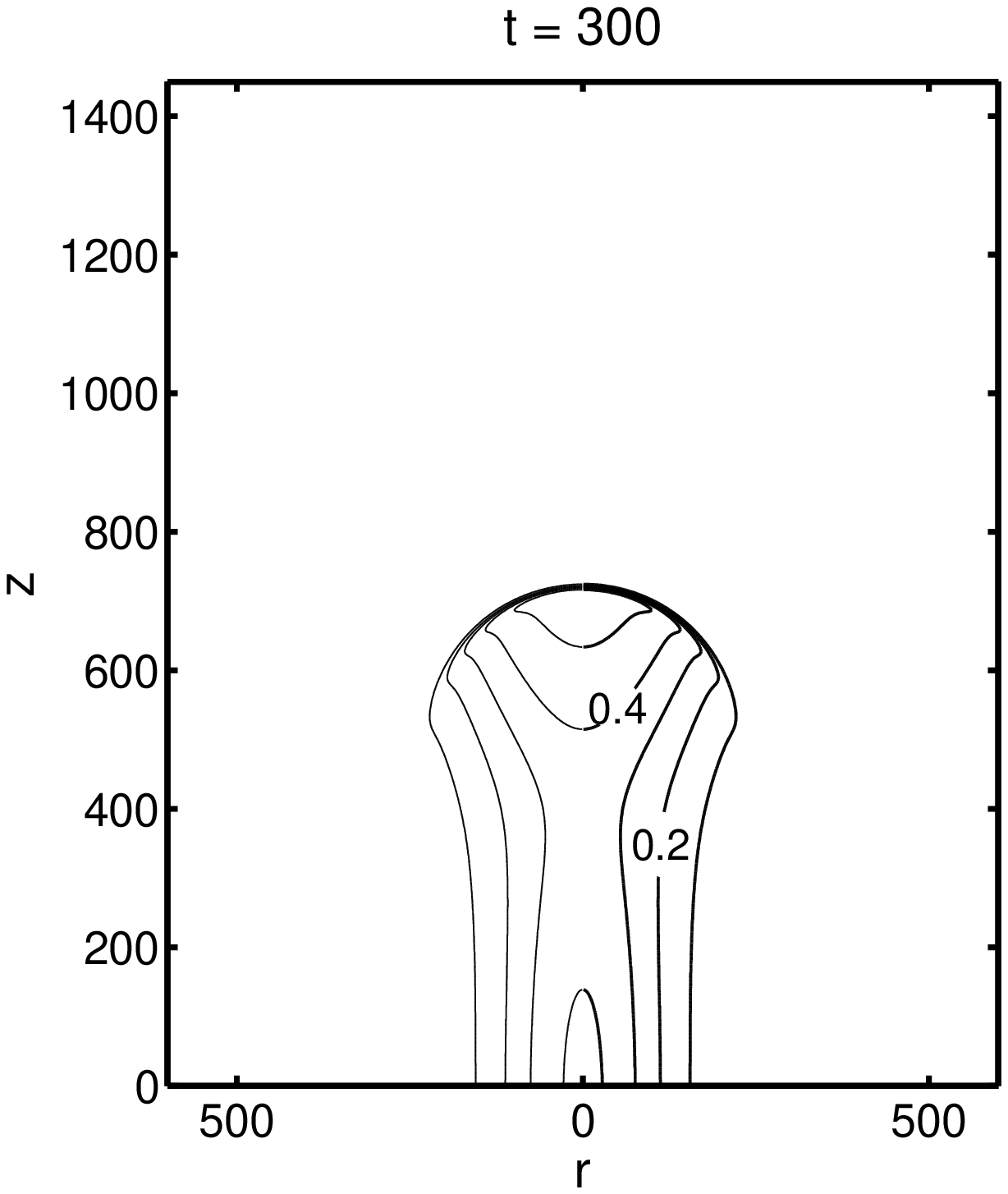}
\includegraphics[width=0.24\linewidth]{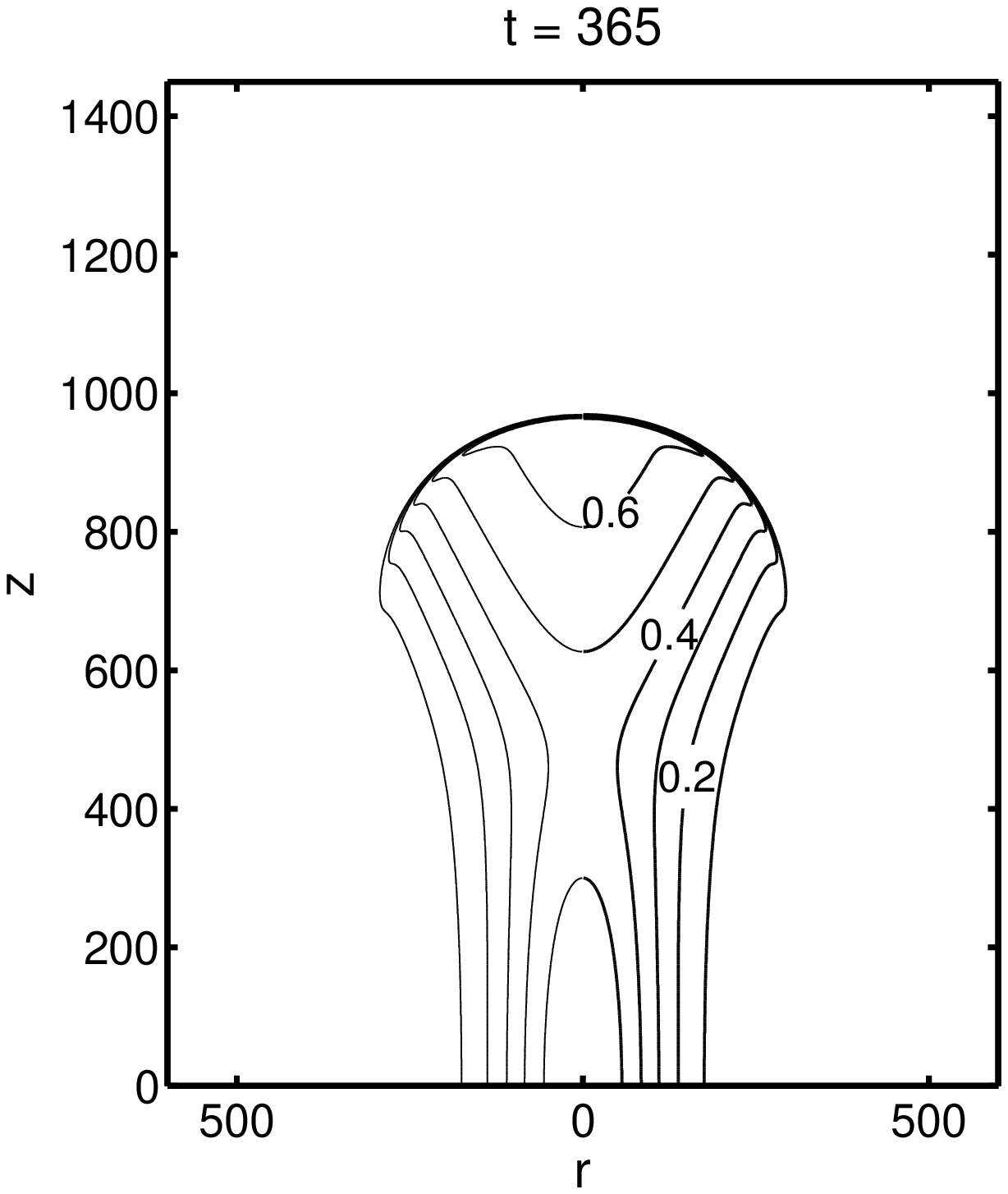}
\includegraphics[width=0.24\linewidth]{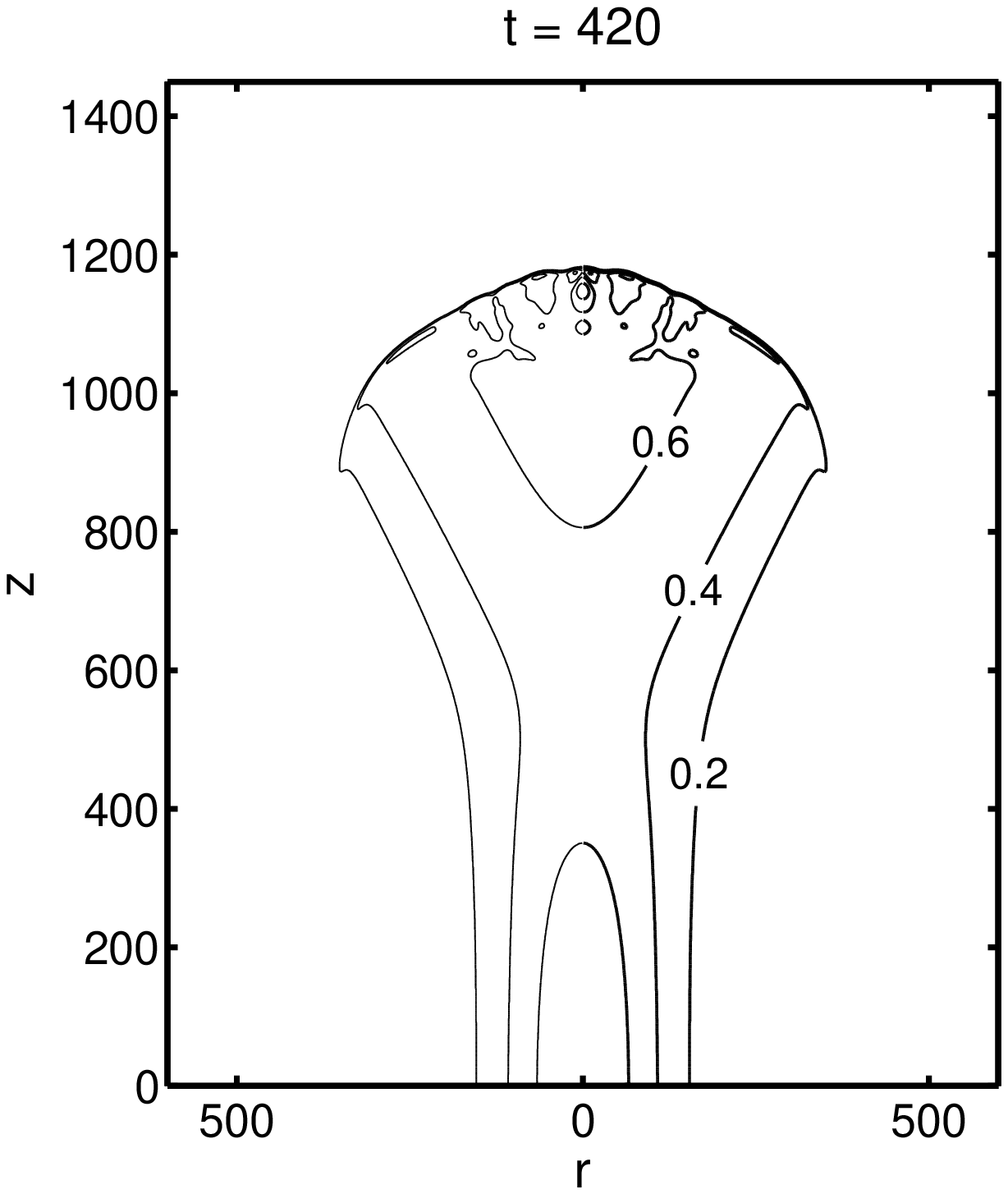}
\includegraphics[width=0.24\linewidth]{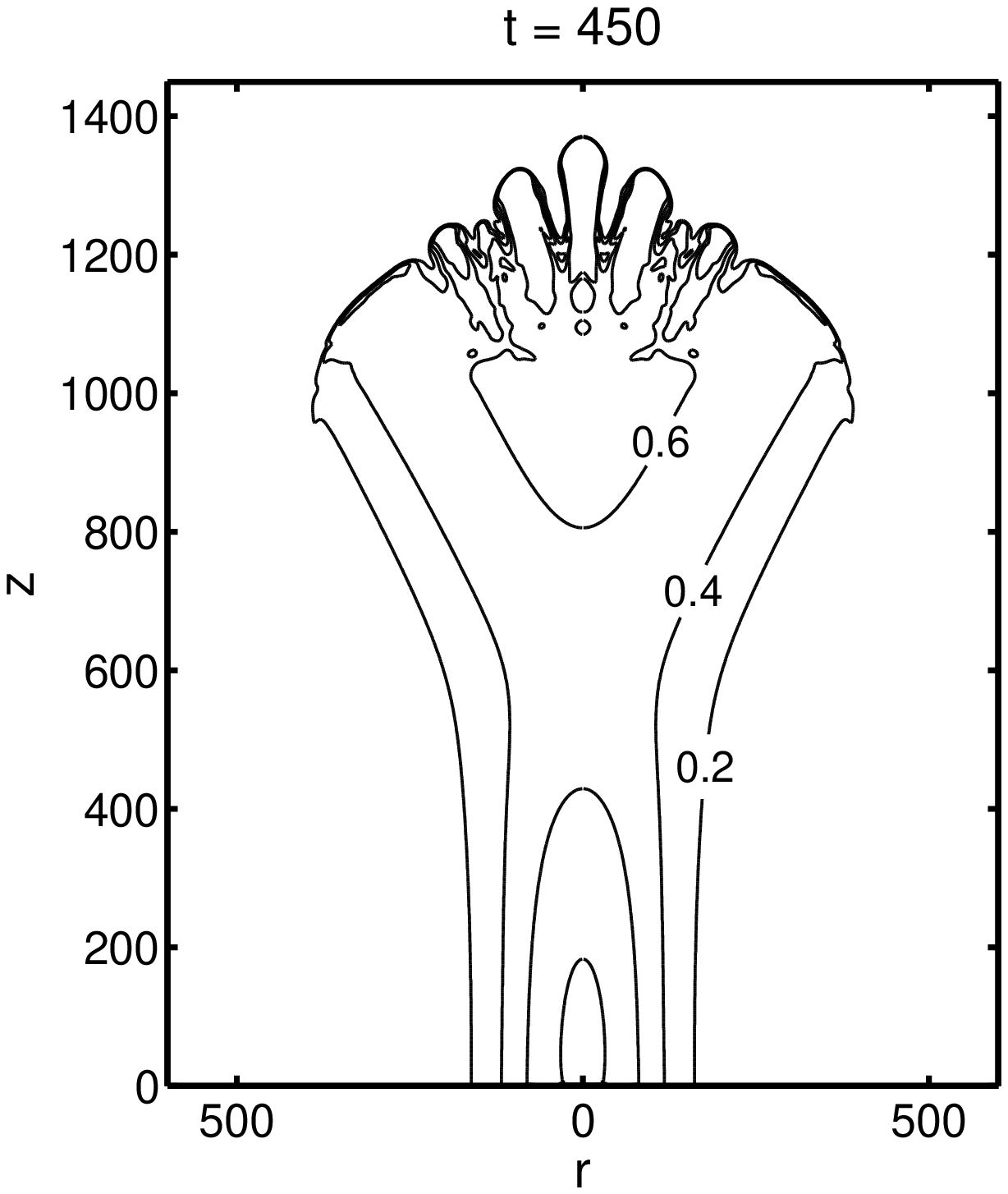}
\caption{Evolution of spontaneous branching of anode directed streamers
in a strong homogeneous background field at times $t=300$, 365, 420 and 450.
Model, initial and boundary conditions are discussed in the text.
The planar cathode is located at $z=0$ and the planar anode at $z=2000$
(shown is $0\le z\le 1400$). The radial coordinate extends from the origin
up to $r=2000$ (shown is $0\le r\le 600$). The lines denote
levels of equal electron density $\sigma$ with increments of  0.2
as indicated by the labels.}
\label{fig:bran}
\end{figure*}

\section{NUMERICAL SIMULATIONS}
In this section we will present details of numerical simulations of the streamer model discussed previously. In confined geometries streamers usually have a nontrivial finger like shape. In general two regions can be observed. The interior of the streamer is an ionized region, quasineutral and equipotential. The outer region is filled with the nonionized gas. Those two regions are separated by a very narrow region in which all the most of the ionization process is taking place. In this same space there is a nonzero charge density and consequently a very large electric field gradient. This is one of the reasons why accurate simulations are rather demanding. These features are strongly reminiscent of what occurs in combustion fronts \cite{Will} and viscous fingering.\cite{Pelce} 

Simulations based on this model as far we know were accomplished by Dhali and Willians \cite{DW} and by Vitello {\it et.~al}.\cite{Vit} There is also some work by Wang and Kunhardt.\cite{Kun} In Fig.~\ref{fig:bran} we can see some simulations of the model. A planar cathode is located at $z=0$ and a planar anode at $z=2000$. The stationary
potential difference between the electrodes $\Delta\Phi=1000$ 
corresponds to a uniform background field ${\bf E} = -0.5 \;{\bf e}_z$
in the $z$ direction. For nitrogen under normal conditions, this corresponds to
an electrode separation of 5 mm and a potential
difference of 50 kV. The unit of time $\tau_0$
is 3 ps, and the unit of field $E_0$ is 200 kV/cm.
We use $D=0.1$ which is appropriate for nitrogen, and assume
cylindrical symmetry. The radial coordinate
extends from the origin up to $r=2000$ to avoid lateral boundary effects on the field configuration. As initial condition, we used an electrically neutral
Gaussian ionization seed on the cathode
\begin{equation}
\label{6}
\sigma(r,z,t=0)=\rho(r,z,t=0)
=10^{-6}\; e^{-(z^2+r^2)/100^2}. 
\end{equation} 

The parameters of our numerical experiment are essentially
the same as in the earlier simulations of Vitello {\it et al.}
\cite{Vit}, except that our background electric field is twice
as high; the earlier work had 25 kV applied over a gap of 5 mm.
This corresponded to a dimensionless background field of 0.25,
and branching was not observed. Further details on this simulation can be found in the references.\cite{ME} 

There have been some more simulations with improved numerics and different boundary and initial conditions \cite{Rocco} where this branching phenomena has been observed.  

\section{The one dimensional streamer equations in a comoving frame: the planar front}
In the previous section some numerical evidence of branching have been presented. In this section and the ones to follow we will investigate this issue by analytical means. Here we will start with the solution for a stationary planar front. The idea is to find a uniformly translating front and investigate how transversal perturbation of this solution will develop.    

For planar fronts, we assume that charge varies only in the $z$ direction, so using the equations (\ref{1})-(\ref{4}) we can write
\begin{eqnarray}
\label{1010}
\partial_t\sigma -\partial_z(\sigma E)&-& D \partial_z^2\sigma
-\sigma f(|E|)=0~,
\nonumber\\
\partial_t\rho&-& \sigma  f(|E|)=0~,
\nonumber\\
\partial_z E&-&\rho+\sigma =0~.
\end{eqnarray} 

Next we will change our reference frame to a frame moving with velocity $v$ in the $z$ direction $(x,y,\xi=z-vt)$. Then equations (\ref{1010}) read
\begin{eqnarray}
\label{1011}
\partial_t\sigma = v\partial_\xi\sigma
&+&\partial_\xi(\sigma E) +D \partial_\xi^2\sigma+\sigma f(|E|),
\nonumber\\
\partial_t\rho = v\partial_\xi \rho &+ & \sigma  f(|E|),
\nonumber\\
\partial_\xi E - \rho&+&\sigma=0~.
\end{eqnarray} 

A front translating uniformly with velocity $v$ in the fixed frame is stationary in this comoving frame, $\partial_t\sigma=\partial_t\rho=0$. As a result, the corresponding front profiles are solutions of ordinary differential equations.

We need to set the boundary conditions. The field, being completely screened
in the ionized region, is approximately constant in space and time
far ahead of the front, so it follows
\begin{equation}
\label{1012}
{\bf E}=\left\{\begin{array}{ll} 0~~&~z\to -\infty\\
                                 E_\infty\;\hat z~~,~E_\infty<0~~&~z\to +\infty
               \end{array} \right.~,
\end{equation}
where $\hat z$ is the unit vector in $z$ direction.
These boundary conditions imply, that a time independent amount of charge
is travelling within the front, and no currents flow far behind
the front in the ionized regime.

Now, for any nonvanishing far field $E_\infty$, there is a continuous
family of uniformly translating front solutions,\cite{Ute,Lagarkov}
since the front propagates into an unstable state.\cite{pulled1}
In particular, for $E_\infty>0$ there is a solution for any velocity
$v\ge0$, and for $E_\infty<0$, there is a solution for any $v\ge|E_\infty|$.
These solutions are associated with an exponentially decaying electron 
density profile: an electron profile that asymptotically for large $\xi$
decays like $\sigma(\xi)\propto e^{-\lambda \xi}$ with $\lambda\ge0$.

It will ``pull'' an ionization front along with the same speed.
(For $E_\infty>0$, the same equation applies for all $\lambda\ge 
f(E_\infty)/E_\infty$, hence for $v\ge0$). For the interested reader we refer him to the bibliography.\cite{pulled1}

Dynamically, the velocity is selected by the initial electron profile.\cite{Ute,pulled1} If initially the electron density strictly vanishes 
beyond a certain point $\xi_0$ (corresponding to $\lambda=\infty$ above)
\begin{equation}
\label{init}
\sigma=0=\rho~~\mbox{for }\xi > \xi_0~~\mbox{at }t=0,
\end{equation}
then this will stay true for all times $t>0$ in a coordinate system
moving with velocity $v=|E_\infty|$,
and an ionization front propagating precisely with the electron 
drift velocity $|E_\infty|$ develops. In the remainder of the paper,
we will consider this particular case. 

\begin{figure}[htbp]
  \centering
\includegraphics[width=0.45\textwidth]{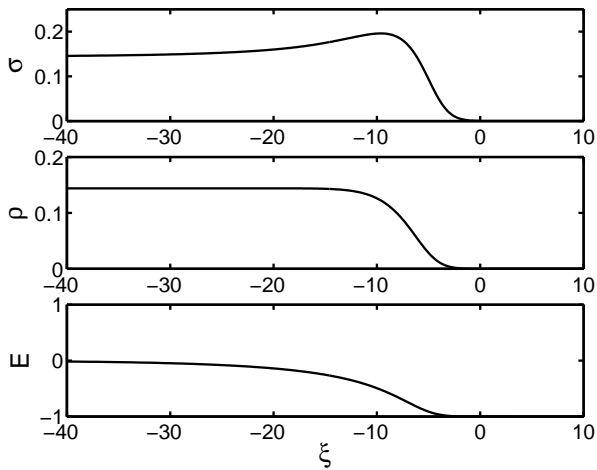}
  \caption{Electron density $\sigma$, ion density $\rho$ and electric field 
$E$ for a negative ionization front moving with 
$v=|E_\infty|$ in the comoving frame and. The far field is $E_\infty=-1$ and $D=0.1$.}
  \label{fig:dsol}
\end{figure}

In Fig.~\ref{fig:dsol} we have solved (\ref{1011}) with the boundary conditions discussed previously (\ref{1012}) and (\ref{init}). We have chosen the far field $E_\infty=-1$ and the diffusion coefficient $D=0.1$. It was done using a shooting method for solving our two point boundary value problem. This technique consists in choosing values for all of the dependent variables at one boundary. These values must be consistent with any boundary conditions for that boundary, but otherwise are arranged to depend on arbitrary free parameters whose values we initially randomly guess. We then integrate the ODEs by initial value methods, arriving at the other boundary. In general, we find discrepancies from the desired boundary values there. Now we adjust the free parameters at the starting point that zeros the discrepancies at the other boundary. The idea is to iterate this procedure until obtaining the desired accuracy. There is a nice pedestrian explanation of solving boundary value problems by shooting in Numerical Recipes.\cite{Num} 

\section{Shock Fronts}
In this section we will simplify a bit more our model by taking the limit $D\to 0$. For negative fronts, the limit $D\to 0$ is smooth and eliminates the algebraic relaxation.\cite{pulled1} It also reduces the order of the equations. We therefore make $D=0$ in the streamer equations. Then, in the comoving frame, using (\ref{1011}) we can write for a stationary front 
\begin{eqnarray}
\label{ab1}
 v\partial_\xi\sigma &+&\partial_\xi(\sigma E)+\sigma f(|E|)=0,\\
\label{ab2}
 v\partial_\xi \rho &+ & \sigma  f(|E|)=0,\\
\label{ab3}
\partial_\xi E &-& \rho+\sigma=0~.
\end{eqnarray} 

We can solve this system analytically. If we take the (\ref{ab2}) and subtract it from (\ref{ab1}), using (\ref{ab3}) to eliminate $\sigma - \rho$, we get
\begin{equation}
\label{205}
-v\partial_\xi E+\sigma E=0.
\end{equation}
This equation is just a consequence of the charge conservation. We can see this by writing $\partial_tq+\nabla\cdot{\bf j_{tot}}=0$, with the total charge defined as $q=\rho-\sigma$. In our model, each ionizing collision, produces the  same number of negative and positive charge, so we end with $\nabla\cdot{\bf j_{tot}}=0$. The total current is given by ${\bf j_{tot}} = \partial_t{\bf E} + \sigma{\bf E}$ and for a planar front with constant and time independent field 
${\bf E}=E_\infty\hat z$ (\ref{1011}) in the non-ionized region where $\sigma=0$, the total current ${\bf j}_{tot}=j_{tot}(t)\hat z$ vanishes. 
In the comoving frame of Eqs.\ (\ref{1011}) and (\ref{ab1})--(\ref{ab3}),
this means (\ref{205}).

The front equations now reduce
to two ordinary differential equations for $\sigma$ and $E$
\begin{eqnarray}
\partial_\xi[(v+E)\sigma]&=&-\sigma f(E)~~,~~f(E)=|E|\alpha(E)~,
\nonumber\\
 v\partial_\xi\ln|E|&=&\sigma,
\end{eqnarray}
that can be solved analytically to give
\begin{eqnarray}
\label{2012}
\sigma[E]&=&\frac{v}{v+E}\;\rho[E],\\
\label{2013}
\rho[E]&=&\int^{|E_\infty|}_{|E|}\!\!\frac{f(x)}{x}\;dx
=\int^{|E_\infty|}_{|E|}\!\!\!\!\!\!\alpha(x)dx,\\
\label{2014}
\xi_2-\xi_1&=&\int_{E(\xi_1)}^{E(\xi_2)}\frac{v+x}{\rho[x]}\;\frac{dx}{x}~.\\
\nonumber
\end{eqnarray}
This gives us $\sigma$ and $\rho$ as functions of $E$, 
and the space dependence $E=E(\xi)$ implicitly as $\xi=\xi(E)$ 
in the last equation.

\begin{figure}[htbp]
  \centering
  \includegraphics[width=0.45\textwidth]{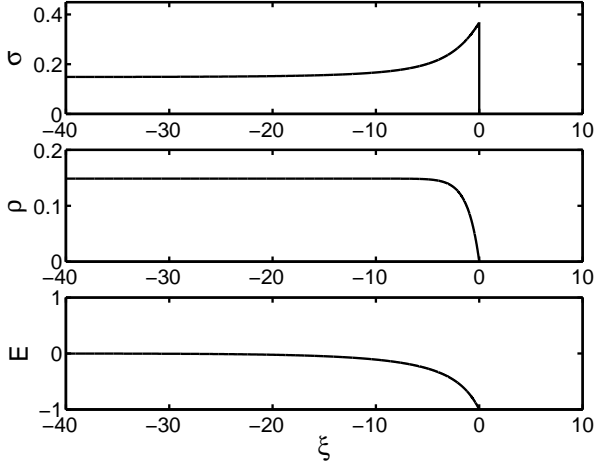}
  \caption{Electron density $\sigma$, ion density $\rho$ and electric field $E$ for a negative ionization shock front moving with $v=|E_\infty|$ in the comoving frame. The far field is $E_\infty=-1$.}
  \label{fig:shock}
\end{figure}

We have plotted in Fig.~\ref{fig:shock} the solutions (\ref{2012})--(\ref{2014}) for a shock front moving with $v=1$. We have chosen $\xi_1=0$ and then $E(\xi_1)=E_\infty$

\section{Study of instabilities: corrugation of the front}
In this section we will study the stability of the planar shock front. It may be unstable with respect to perturbations having periodicity on the surface of discontinuity and then forming ``ripples'' or ``corrugations'' on that surface. In that case, we will be interested in obtaining the dispersion relation curve to find which mode will grow faster and eventually determine the streamer characteristic shape. Here we will derive the perturbed equations and the boundary conditions. 

Let the planar shock front which propagates into the $z$ direction receive a slight perturbation having an arbitrary dependence on the transversal coordinates $x$ and $y$. Within linear perturbation theory, they can be decomposed into
Fourier modes. Therefore we need the growth rate $s(k)$ of an arbitrary 
transversal Fourier mode to predict the evolution of an
arbitrary perturbation. Because of isotropy within the transversal 
$(x,y)$-plane, we can restrict the analysis to Fourier modes in the 
$x$ direction, so we study linear perturbations $\propto \exp(st+ikx)$. The notation anticipates the exponential growth of such modes. Any perturbation will also lead to a perturbation of the position of the ionization shock front. So we will introduce the new variable $\zeta=\xi-\epsilon\;\exp(ikx+st)$ and the ansatz
\begin{eqnarray}
\label{pertu}
\sigma(x,\zeta,t)&=&\sigma_0(\zeta)+\epsilon\;\sigma_1(\zeta)\;e^{ikx+st},
\nonumber\\
\rho(x,\zeta,t)&=&\rho_0(\zeta)+\epsilon\;\rho_1(\zeta)\;e^{ikx+st},
\nonumber\\
\phi(x,\zeta,t)&=&\phi_0(\zeta)+\epsilon\;\phi_1(\zeta)\;e^{ikx+st},
\end{eqnarray}
where $\sigma_0$, $\rho_0$ and $\phi_0$ are the electron density, 
ion density and electric potential of the planar ionization shock
front obtained in the previous section. Note, however, that these planar
solutions are shifted to the position of the perturbed front.
Substitution of these expressions into (\ref{1011}) (with $D=0$) gives to leading order in the small parameter $\epsilon$
\begin{eqnarray}
\label{EQ}
(v+E_0)\;\partial_{\zeta}\sigma_1&=&(s+2\sigma_0-\rho_0-f)\;\sigma_1
\nonumber\\
& &-\sigma_0\;\rho_1+(\partial_{\zeta}\sigma_0-\sigma_0 f')\;
\partial_{\zeta} \phi_1-s\partial_{\zeta}\sigma_0,
\nonumber\\
v\;\partial_{\zeta}\rho_1&=&-f\;\sigma_1
+s\;\rho_1-\sigma_0 f'\;\partial_{\zeta} \phi_1-s\partial_{\zeta}\rho_0,
\nonumber\\
\left(\partial_{\zeta}^2-k^2\right)\;\phi_1&=&\sigma_1-\rho_1+k^2E_0.
\end{eqnarray}
In equations (\ref{EQ}) we denote $f=f(E_0)$, $f'=\partial_{|E|}f(|E|)\Big|_{E_0}$, and 
$E_0 = -\partial_{\zeta} \phi_0(\zeta)$ as the electric field of the uniformly translating front. In the third equation, the term $k^2\phi_1$ comes as a consequence of the dependence of the electric potential with $x$. 

These equations can be written in matrix form as 

\begin{eqnarray}
\label{Matr1}
&&\partial_{\zeta} 
\left(\begin{array}{c}\sigma_1\\ \rho_1\\ \psi_1\\ \phi_1\end{array}\right)
={\bf M}_{s,k}\cdot
\left(\begin{array}{c}\sigma_1\\ \rho_1\\ \psi_1\\ \phi_1\end{array}\right)
- \left(\begin{array}{c}s\partial_{\zeta}\sigma_0/(v+E)\\ s\partial_{\zeta}\rho_0/v\\ - Ek^2\\0 
        \end{array}\right)
\nonumber\\
&&\\
\label{Matr2}
&&{\bf M}_{s,k}=\left(\begin{array}{cccc}
  \displaystyle\frac{s+2\sigma_0-f-\rho_0}{v+E}&\displaystyle\frac{-\sigma_0}{v+E}&
      \displaystyle\frac{\partial_{\zeta}\sigma_0-\sigma_0 f'}{v+E}&0\\ 
  &&&\\
  \displaystyle\frac{-f}{v}&\displaystyle\frac{s}{v}&
      \displaystyle\frac{-\sigma_0 f'}{v}& 0\\ 
  &&&\\
  \displaystyle1&\displaystyle-1&\displaystyle0&k^2\\
  &&&\\
  0&0&1&0
\end{array}\right)
\nonumber\\
\end{eqnarray} 
Note we have introduce an auxiliary field $\psi_1 = \partial_{\zeta} \phi_1$
which coincides with the correction for the electric field sign reversed to order $\epsilon$.

Having obtained the linear order perturbation equations, we are now in position to discuss boundary conditions. First we consider the boundary conditions at $\zeta=0$. There are two types of boundary conditions, some arising
from the boundedness of densities to the left of the shock front
at $\zeta\uparrow0$, and some arising from the continuity of fields 
across the position $\zeta=0$ of the shock front.
From (\ref{ab1}) we gather that $(v+E)\;\partial_{z}\sigma$
is finite for all $z$, also for $z\uparrow0$ and for $z=0$,
since $(v+E)\;\partial_{z}\sigma=\sigma\;(\sigma-\rho-f)$
is finite. The same is true for $\sigma_0$. 
In particular, $\int_{-l}^{l}dz\;(v+E)\;
\partial_{z}\sigma_0\to0$ as $l\to0$, and
$(v+E)\;\partial_{z}\sigma_0\to0$ as $z\uparrow0$.

Therefore we impose the same conditions for $\sigma_1$, namely
\begin{eqnarray}
\label{3012}
\lim_{l\to0}\int_{-l}^{l}d\zeta\;(v+E)\;\partial_{\zeta}\sigma_1=0\\
\label{3013}
\lim_{\zeta\to0^-}\;(v+E)\;\partial_{\zeta}\sigma_1=0
\end{eqnarray}

In a second step we are going to make use of the continuity conditions. We match the $\zeta<0$ solution to the $\zeta>0$ solution. As in front of the shock there are not sources, one has to solve $\nabla^2\phi=0$ for $\zeta>0$ and $\nabla \phi = -E_\infty\;\hat z = v ~\hat z$ when $\zeta\to\infty$. The solution to first order in $\epsilon$ has the form
\begin{equation}
\label{zg0}
\begin{array}{ccl}\sigma&=&0\\
\rho&=&0 \hspace{2cm} \\
\phi&=&a+v\zeta+\epsilon(v+b\;e^{-k\zeta})\;e^{ikx+st}\end{array}
~~~\mbox{for }\zeta>0
\end{equation}
with the undetermined integration constants $a$ and $b$.

Now $\rho$ and $\nabla\phi$ have to be continuous across 
the shock front: $\nabla\phi$ is continuous because the charge density $\rho-\sigma$ is finite everywhere. The continuity of $\rho$ we get from (\ref{ab2}) and the fact, that $\sigma$ and $|{\bf E}|$ are bounded for all $z$.

From the continuity of $\rho$ turns out
\begin{eqnarray}
\label{bc1}
\lim_{\zeta\to0}\Big(\rho(x,\zeta^+,t)-\rho(x,\zeta^-,t)\Big)&=&0
\nonumber\\
~~\Rightarrow~~
\rho_1(0)&=&0
\end{eqnarray}
where we have use (\ref{zg0}) and (\ref{pertu}) to the right and left limits. 

The continuity of the electric field to first order in $\epsilon$ implies that 
\begin{eqnarray}
\lim_{\zeta\to0}\Big(\partial_{\zeta}\phi(x,\zeta,t)|_{\zeta^+}
                       -\partial_{\zeta}\phi(x,\zeta,t)|_{\zeta^-}\Big)=0
\nonumber\\
\lim_{\zeta\to0}\Big(\partial_{x} \phi(x,\zeta,t)|_{\zeta^+}
  -\partial_{x} \phi(x,\zeta,t)|_{\zeta-}\Big)=0
\end{eqnarray}

Using expressions (\ref{zg0}) and (\ref{pertu}) again, these conditions turn out
\begin{eqnarray}
\label{bc2}
\psi_1(0)=-kb \;\;\;\;,\;\;\;\; \phi_1(0)=v+b
\end{eqnarray}

If we impose the continuity of the potential, we get $a=\phi_0(0)$ and $\phi_1(0)=v+b$ (which is the same condition obtained from the continuity of the electric field). 

Finally, from (\ref{3012}) and (\ref{3013}), and taking into account that $f = \sigma_0$ when $\zeta\to0$, we have

\begin{eqnarray}
\label{bc3}
\psi_1(0)=s   \;\;\;\;,\;\;\;\; \sigma_1(0) = \frac{sff'}{s+f}
\end{eqnarray}

Collecting all the identities (\ref{bc1}), (\ref{bc2}) and (\ref{bc3}) we get for the limit of $\zeta\uparrow0$
\begin{equation}
\label{Init}
\left(\begin{array}{c}\sigma_1\\ \rho_1\\ \psi_1\\ \phi_1\end{array}\right)
\stackrel{z\uparrow0}{\longrightarrow}
\left(\begin{array}{c}sf'(v)/(1+s/f(v))\\ 0\\ s\\ (vk-s)/k\end{array}\right)
\end{equation}

The other boundary conditions, at $\zeta=-\infty$ are the total charge equals to zero and the electric field vanishes, so they read

\begin{eqnarray}
\label{End}
\left(\begin{array}{c}\sigma_1\\ \rho_1\\ \psi_1\\ \phi_1\end{array}\right)
\stackrel{\zeta\downarrow -\infty}{\longrightarrow}
\left(\begin{array}{c}\sigma_1^-\\ \sigma_1^-\\ 0\\ \phi_1^-\end{array}\right)
\end{eqnarray}
where $\sigma_1^-$ and $\phi_1^-$ are constants.

\section{Dispersion Curve}
In the preceding section we have formulated an eigenvalue problem. Given $k$, we want to find $s(k)$ such that we can find a solution for the transversal perturbation equations (\ref{Matr1}) fulfilling the boundary conditions derived previously (\ref{Init}) and (\ref{End}). In general, an analytic treatment for any value of $k$ is not possible and one has to resort to numerical calculations.\cite{PRE} However, in the limits of small and large wave number the equations simplify and we can obtain the asymptotic behaviour of the dispersion relation $s(k)$.

We will start looking at the small $k$-limit. If expressions  
 (\ref{Matr1}) and (\ref{Matr2}) are evaluated only up to first order in $k$, then $\phi_1$ decouples, and we get
\begin{equation}
\label{3031}
\partial_{\zeta} \left(\begin{array}{c}\sigma_1\\ \rho_1\\ \psi_1\end{array}\right)
={\bf N}_{s,k}\cdot
\left(\begin{array}{c}\sigma_1\\ \rho_1\\ \psi_1\end{array}\right)
- \left(\begin{array}{c}\partial_{\zeta}\sigma/(v+E)\\ \partial_{\zeta}\rho/v\\ 0 
        \end{array}\right)+O(k^2)~,
\end{equation}
where 
\begin{equation}
\label{3032}
{\bf N}_{s,k}
=\left(\begin{array}{ccc}
  \displaystyle\frac{s+2\sigma-f-\rho}{v+E}&\displaystyle\frac{-\sigma}{v+E}&
      \displaystyle\frac{\partial_{\zeta}\sigma-\sigma f'}{v+E}\\ 
  &&\\
  \displaystyle\frac{-f}{v}&\displaystyle\frac{s}{v}&
      \displaystyle\frac{-\sigma f'}{v}\\ 
  &&\\
  \displaystyle1&\displaystyle-1&\displaystyle0\\
\end{array}\right)+O(k^2)
\end{equation}
is the truncated matrix ${\bf M}_{s,k}$ (\ref{Matr2}).
The fourth decoupled equation reads
\begin{equation}
\label{3033}
\partial_{\zeta}\phi_1=\psi_1
\end{equation}

The boundary condition (\ref{Init}) turns out 
\begin{equation}
\label{3034}
\left(\begin{array}{c}\sigma_1\\ \rho_1\\ \psi_1\end{array}\right)
\stackrel{{\zeta}\uparrow0}{\longrightarrow}
\left(\begin{array}{c}
f'/(1+s/f)\\ 0 \\ 1\end{array}\right)+O(k^2)
\end{equation}
and 
\begin{equation}
\label{3035}
\phi_1(0)=\frac{vk-s}{sk}=\frac{v}{s}-\frac{1}{k}
\end{equation}
\\
\\
The expressions (\ref{3033}) and (\ref{3035}) give a condition on $\psi_1$
\begin{equation}
\label{3036}
\frac{vk-s}{sk}=\int_{-\infty}^0 \psi_1({\zeta})\;d{\zeta}.
\end{equation}

Consider now the limit $s\ll f(v)$. Then Eqs.\ (\ref{3031})
and (\ref{3034}) up to order $s/f(v)$ become identical 
to the perturbed equations obtained from an infinitesimal change of 
$E_\infty$. If we compare two uniformly translating fronts with infinitesimally different field $E_\infty$ at identical positions, their linearised difference solves the same equations.  In this case, $\psi_1$ is 
independent of $s$ and $k$. But then (\ref{3036}) implies
\begin{equation}
s=vk+O(k^2)~~~\mbox{for }~~k\ll\alpha(v)~.
\end{equation}

This result also has an immediate physical interpretation:
$1/k$ is the largest length scale involved. It is much larger
than the thickness of the screening charge layer. Therefore
the charge layer can be contracted to a $\delta$-function
contribution along an interface line.
Such a screening charged interface precisely has the
instability mode $s=vk$.

In the opposite limit, when $k$ becomes large enough, we can also find a relation for the dispersion curve. We will need to make the assumption that the ion and electron densities remain bounded. Taking this into account, we can write 
using (\ref{Matr1}) the  equations for $\psi_1$ and $\phi_1$ as
\begin{eqnarray}
\label{bigk}
\partial_{\zeta} \psi_1  &\simeq& k^2 \left(\phi_1+E \right), \nonumber\\
\partial_{\zeta} \phi_1 &=& \psi_1
\end{eqnarray}

On the short length scale $1/k$, the unperturbed electric field 
for $\zeta<0$ can be approximated making an asymptotic expansion of (\ref{2012})--(\ref{2014}) by \cite{PRE}
\begin{eqnarray}
E \simeq -v -f(v)\zeta,
\end{eqnarray}

Inserting this expression in (\ref{bigk}), we obtain
\begin{eqnarray}
\label{phi1}
\partial_\zeta^2\phi_1=k^2\Big(\phi_1-v-f(v)\zeta\Big).
\end{eqnarray}
The boundary condition (\ref{Init}) fixes $\phi_1(0)=(vk-s)/k$ and 
$\psi_1(0)=\partial_\zeta\phi_1=s$. The unique solution of (\ref{phi1})
with these initial conditions is
\begin{eqnarray}
\phi_1(\zeta) = v + f(v)\zeta - \frac{f(v)}{2k}\; e^{k\zeta} 
+ \frac{f(v)-2s}{2k}\; e^{-k\zeta}
\end{eqnarray}
for $\zeta<0$. Now the mode $e^{-k\zeta}$ would increase rapidly 
towards decreasing $\zeta$, create diverging electric fields 
in the ionized region and could not be balanced by any other terms 
in the equations. Therefore it has to be absent. The demand that its 
coefficient $(f(v)-2s)/2k$ vanishes, fixes the dispersion relation
\begin{eqnarray}
\label{s=f/2}
s(k)=\frac{f(v)}{2}+O\left(\frac1{k}\right)~~~\mbox{for }~~k\gg\alpha(v)~.
\end{eqnarray}

Again there is a simple physical interpretation of this growth rate.
The electric field can be approximated in leading order by
\begin{eqnarray}
{\bf E}(x,\zeta,t) \simeq
\left\{\begin{array}{ll}-\hat z\;\big(v+f(v)\zeta\big) &\mbox{for }~\zeta<0\\
-\hat z\;v~~&\mbox{for }~\zeta>0\end {array}\right. 
\end{eqnarray}
When the discontinuity propagates with the local field $v=-E$,
a perturbation in a field 
${\bf E}=-\hat z \big(v+\partial_\zeta E\;\zeta\big)$ will grow
with rate $\partial_\zeta E$. The averaged slope of the field 
for $\zeta>0$ and $\zeta<0$ is $\partial_\zeta E=f(v)/2$,
and this is precisely the growth rate (\ref{s=f/2}) determined above.

We have studied the (in)stability of planar negative ionization fronts
against linear perturbations and we have found 
\begin{eqnarray}
\label{sk}
s(k)&=&\left\{\begin{array}{ll}
|E_\infty| \;k &~~\mbox{ for }k\ll\alpha(|E_\infty|)\\
|E_\infty|\;\alpha(|E_\infty|)/2 &~~\mbox{ for }k\gg\alpha(|E_\infty|)
\end{array}\right.
\end{eqnarray}
\\
So the planar front becomes unstable with a linear growth rate $s(k)$ for small $k$ to a saturation value $|E_\infty|\;\alpha(|E_\infty|)/2$. This gives us a mechanism for branching. In the case of a curved front, if the radius of curvature increases, the planar approximation for the tip is sensible and allows a qualitative understanding of the branching phenomena. 

\section{Summary and outlook}
In this paper a fully deterministic model for streamers, without photoionization, which is suitable for nonattaching gases like nitrogen has been presented. We have proposed that an anode directed front can branch spontaneously according to this model due to Laplacian interfacial instability. We have shown some numerical evidence of this phenomena. We have studied the stability of a planar front and how transversal perturbation would grow. This gives us a qualitative picture of the mechanism acting on a curved front, and we have got the asymptotic behaviour of the dispersion curve. 

However, some questions remain to be answered. From the dispersion curve any short enough wave length instability will grow. We do expect that a regularization mechanism should come into play. This regularization mechanism which selects a particular mode could be the electric screening due to curvature. In the present it is under investigation.\cite{PRE} Other possibility could be the diffusion phenomena not considered in the shock front case. Diffusion was neglected to prevent mathematical challenges, but soon or later one has to face challenges. 

In any case, the physics of low temperature plasmas is an area where many fundamental questions are still open, where ideas from patter formation, electrodynamics, quantum mechanics, statistical mechanics and nonlinear mathematics can be applied, and where the experimental side has been ahead of the theoretical one. When I hear some pessimistic voices for the future of physics, I always think there is much room at the bottom...

\end{document}